\documentclass{article}

      \PassOptionsToPackage{compress,sort}{natbib}

     \usepackage[final]{neurips_2020_ml4ps}

\usepackage[utf8]{inputenc} 
\usepackage[T1]{fontenc}    
\usepackage{hyperref}       
\usepackage{url}            
\usepackage{booktabs}       
\usepackage{amsfonts}       
\usepackage{nicefrac}       
\usepackage{microtype}      
\usepackage{amsmath}
 
\usepackage{caption}
\usepackage{subcaption}
\usepackage{graphicx}
\usepackage{xcolor,listings}
\graphicspath{ {./images/} }

\title{Anomaly Detection in Astronomical Images with Generative Adversarial Networks}

\author{%
    {Kate Storey-Fisher}\\
    Center for Cosmology and Particle Physics,\\
    Dept. of Physics, New York University,\\
    New York, NY, USA\\
    \texttt{k.sf@nyu.edu}\\
    \And
    Marc Huertas-Company \\
    LERMA, Observatoire de Paris, PSL, Paris, France;\\
    CNRS, Sorbonne Universit\'es,
    UPMC Univ. Paris, Paris, France\\
    Departamento de Astrof\'isica, Universidad de La Laguna, Tenerife, Spain\\
    Instituto de Astrof\'isica de Canarias, Tenerife, Spain\\
    Dept. of Physics, UC Santa Cruz, Santa Cruz, CA, USA\\
    \texttt{marc.huertas.company@gmail.com} \\
    \And
    Nesar Ramachandra \\
    High Energy Physics Division,\\ Argonne National Laboratory, \\
    Lemont, IL, USA\\
    \texttt{nramachandra@anl.gov} \\
    \And
    Francois Lanusse \\
    AIM, CEA, CNRS, Universit\'e\\Paris-Saclay, Universit\'e Paris Diderot, \\
    Paris, France\\
    \texttt{francois.lanusse@cea.fr}\\
    \AND
    Alexie Leauthaud \\
    Dept. of Astronomy, \\ UC Santa Cruz,\\
    Santa Cruz, CA, USA  \\
    \texttt{alexie@ucsc.edu}\\
    \And
    Yifei Luo \\
    Dept. of Astronomy, \\ UC Santa Cruz,\\
    Santa Cruz, CA, USA  \\
    \texttt{yluo42@ucsc.edu}
    \And
    Song Huang \\
    Dept. of Astrophysical Sciences,\\
    Princeton University,\\
    Princeton, NJ, USA\\
    \texttt{sh19@astro.princeton.edu}
}

\begin{document}

\maketitle

\begin{abstract}
    We present an anomaly detection method using Wasserstein generative adversarial networks (WGANs) on optical galaxy images from the wide-field survey conducted with the Hyper Suprime-Cam (HSC) on the Subaru Telescope in Hawai'i.\footnote{\url{https://hsc.mtk.nao.ac.jp/ssp/}}
    The WGAN is trained on the entire sample, and learns to generate realistic HSC-like images that follow the distribution of the training data.
    We identify images which are less well-represented in the generator's latent space, and which the discriminator flags as less realistic; these are thus anomalous with respect to the rest of the data.
    We propose a new approach to characterize these anomalies based on a convolutional autoencoder (CAE) to reduce the dimensionality of the residual differences between the real and WGAN-reconstructed images.
    We construct a subsample of $\sim$9,000 highly anomalous images from our nearly million object sample, and further identify interesting anomalies within these; these include galaxy mergers, tidal features, and extreme star-forming galaxies.
    The proposed approach could boost unsupervised discovery in the era of big data astrophysics.
\end{abstract}

\section{Introduction}

Many discoveries in astronomy have been made by identifying unexpected outliers in collected data (e.g. \citealt{Cardamone2009}, \citealt{Massey2019}). 
As data sets increase in size, automated methods for detecting these outliers are critical; for example, the upcoming Rubin Observatory will observe 40 billion objects \citep{Ivezic2018}.
This and other large surveys present opportunities for novel discoveries in their massive data sets.

Unsupervised machine learning lends itself to this problem, as it allows for outlier identification without expert labelling or assumptions about expected outliers.
These methods have already proven useful in astronomy (e.g. \citealt{Baron2017,Solarz2017,Giles2019,Ishida2019,Pruzhinskaya2019,Lochner2020}).
Generative adversarial networks (GANs) have a natural application to identifying outliers, as they are able to model complex distributions of high-dimensional data.
The generator better models images that are more common in the training set, and more poorly models images that are rare or anomalous relative to the rest of the data.
The discriminator distinguishes between real and generated images, and identifies these poor generator reconstructions as they are less realistic.
The discriminator has proven particularly useful for outlier detection (e.g. \citealt{Son2019}), making GANs a worthwhile alternative to simpler deep learning approaches such as autoencoders or variational autoencoders (VAEs, \citealt{Kingma2014}). 
GANs were first applied to anomaly detection by \cite{Schlegl2017}, in the context of medical imaging, and have since been used to detect outliers in time-series data \citep{Li2018}.
Recently \cite{Margalef-Bentabol2020} used a GAN to detect merging galaxies and compare galaxy simulations against observations; they show that this outperforms traditional k-means clustering.

In this work, we train a Wasserstein GAN to identify anomalous objects in a subsample of images of the deep sky.
We then characterize the anomalous images with a new convolutional autoencoder-based approach, and identify a set of scientifically interesting anomalies.

\section{Data}

We use data from the Hyper Suprime-Cam Subaru Strategic Program (HSC-SSP).
The wide-field optical survey is imaged with the Subaru Telescope in 5 filters.
The second public data release (PDR2, \citealt{Aihara2014}) contains over 430 million primary objects in the wide field covering 1114 deg$^2$.

We start from a catalog of detected objects and choose a magnitude slice for our analysis ($20.0<i<20.5$) for a consistent sample in object size.
(This simplifies the analysis, as a wide range of magnitudes might bias the detector towards identifying all bright objects as anomalies.
For larger samples, one could train separate WGANs on each slice, or carefully choose training batches to balance magnitudes; we leave this for future work.)
We generate cutouts of $96 \times 96$ pixels ($\sim$15$\times$15 arcsec); this captures the entirety of most objects while still being a reasonable size for the network.
This results in a sample of 942,782 objects, consisting of $\sim$70\% extended and $\sim$30\% compact objects.

We first preprocess the images to avoid issues due to the raw data range spanning multiple orders of magnitude.
We convert the $gri$-band fluxes to RGB values using the method of \citealt{Lupton2004}, producing values from 0 to 255 for each pixel in each band, and then normalize these values individually to between 0 and 1.
While this may affect the features identified as anomalous, such as suppressing extreme fluxes, the rescaled images largely retain the information about each object.\

\section{Model \& Training}

\subsection{WGAN Architecture and Training}

We construct a standard Wasserstein generative adversarial network with Gradient Penalty based on the implementation by \cite{Gulrajani2017}.
We construct our generator to have a depth of 4 with a latent space of dimension 128, and a sigmoid activation function.
The discriminator also has a depth of 4.
We train the WGAN in batches of 32 images, with 5 discriminator updates per generator update.
We finalize the model at 10,000 training iterations, after which the generator and discriminator losses stabilize.
We note that the use of the Wasserstein distance helps protect against convergence issues common of GANs.
We also perform checks that the WGAN has not fallen into various failure modes, including mode collapse and domination by either the generator or discriminator.

\subsection{Anomaly Score Assignment}

The basic procedure for anomaly detection involves setting the WGAN to generate its best reconstruction of each image.
A poorly reconstructed image should indicate an object that is anomalous with respect to the rest of the sample, as the WGAN has learned the global distribution of the data and will be better at generating ``typical'' images.
This approach requires an inverse mapping from images to the WGAN's latent space and a quantification of anomalousness.

To perform this inverse mapping, we use the trained generator and discriminator from the WGAN (no longer updating the weights).
Typically one starts from a random draw from the WGAN's latent space, and optimizes to find the vector that achieves the best reconstruction of a given image (e.g. \citealt{Schlegl2017}).
Specifically, we aim to minimize a loss $\mathcal{L}$ based on the difference between the original and reconstructed image, in both pixel-space and feature-space.
The generator score $s_\mathrm{gen}$ is the mean square error (MSE) of the pixel-wise difference between the generator-reconstructed image and the original.
We also use the discriminator to perform feature-matching to capture whether the reconstruction contains similar features as the original.
The discriminator score $s_\mathrm{disc}$ is calculated by feeding this generator reconstruction through the discriminator and extracting its 6x6 representation of the input from the penultimate layer, doing the same for the original image, and computing the MSE between these.
The total loss is then $\mathcal{L} = (1-\lambda) \, s_\mathrm{gen} + \lambda \, s_\mathrm{disc}$, where $\lambda$ is a weighting hyperparameter.
For this application we choose $\lambda=0.05$ to balance the difference in raw score distributions, though this choice does not have a significant effect as the scores are highly correlated.

This minimization procedure is a time-limiting step for large samples, so we propose an improvement: we first train an encoder, a straightforward convolutional network, on the entire training sample to make a first approximation of the latent-space vector.
This encoder simply provides a better initial guess of the latent-space location and does not significantly affect the final reconstruction.
We then start from the encoder approximation and, for each image individually, perform a basic minimization of $\mathcal{L}$, optimizing for 10 iterations (though the score usually converges before this).
This final loss value $\mathcal{L}_\mathrm{final}$ quantifies the degree of anomaly of the image, so we assign this to be the image's anomaly score, $s_\mathrm{anom} = \mathcal{L}_\mathrm{final}$. 
Higher anomaly scores indicate more anomalous objects, while lower scores indicate more typical objects; the scores are relative and meaningful only with respect to the rest of the sample.

\begin{figure}[t!]
\begin{subfigure}{.325\textwidth}
  \centering
  \includegraphics[width=1\linewidth]{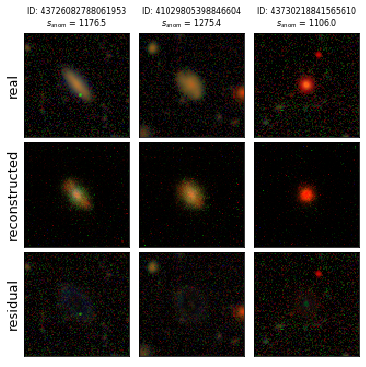}  
  \caption{}
  \label{fig:recon_neg}
\end{subfigure}
\hfill
\begin{subfigure}{.325\textwidth}
  \centering
  \includegraphics[width=1\linewidth]{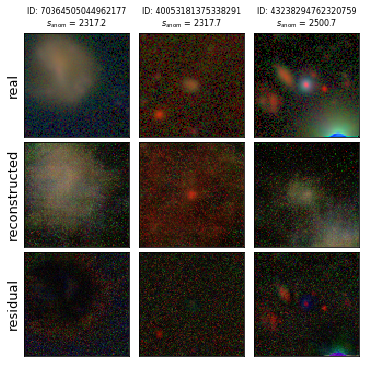}  
  \caption{}
  \label{fig:recon_3sig}
\end{subfigure}
\hfill
\begin{subfigure}{.325\textwidth}
  \centering
  \includegraphics[width=1\linewidth]{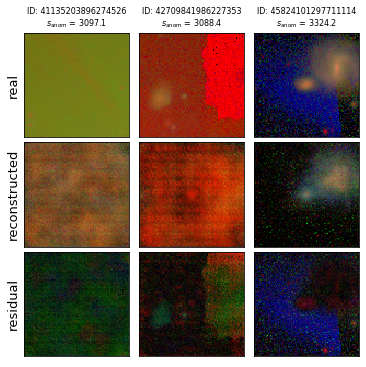}
  \caption{}
  \label{fig:recon_5sig}
\end{subfigure}
\caption{The results of WGAN image reconstruction. The top row of each panel shows the original image, the second row shows the best WGAN reconstruction, and the bottom row shows the residual between the two. The assigned anomaly score is shown at the top of each column. The images in each panel are random samples of images in the following ranges of anomaly score: (a) significantly below the mean, (b) around $3\sigma$ above the mean, (c) greater than $5\sigma$ above the mean. It is clear that higher anomaly scores are indicative of poorer WGAN reconstructions and hence larger residuals.}
\label{fig:recon}
\end{figure}

The result of this process is shown in Figure \ref{fig:recon}.
We can see that the WGAN is able to generate realistic images; for compact objects with standard colors, it constructs images nearly identical to the original, and assigns the objects low anomaly scores (Figure \ref{fig:recon_neg}).
The model is more challenged to generate objects with rare features or colors, as in the images with scores around $3\sigma$ above the mean shown in Figure \ref{fig:recon_3sig}.
Finally, objects with pipeline errors, such as complete saturation in one of the color bands, have extremely high anomaly scores, as the WGAN struggles to reconstruct them (Figure \ref{fig:recon_5sig}).

We identify 9,648 objects with scores greater than 3$\sigma$ above the mean, just over 1\% of the data set; we take these as our high-anomaly sample and perform further characterization on these objects.

\section{Characterization of Anomalies}

\subsection{Dimensionality Reduction with a Convolutional Autoencoder}

A general problem with anomaly detection is distinguishing potentially interesting objects from uninteresting data issues.  
We propose here a new approach based on convolutional autoencoders (CAEs) to postprocess and explore identified anomalies. 
We expect the residual images, the difference between the real and reconstructed images, to contain information about why the WGAN marked an object as anomalous.
(We use the absolute difference as the RBG scale is restricted to positive pixel values, though this does lose potentially useful information.)
However, the pixel space is very high-dimensional, and contains information less relevant to the anomalous features we are interested in, such as background noise.
To reduce the dimensionality of the data and isolate the relevant information, we train a CAE to map the pixels to a 32-dimensional vector.
The CAE has 4 encoding and 4 decoding layers, and uses a standard MSE loss between the true and reconstructed image.

\begin{figure}[t!]
\begin{subfigure}{.495\textwidth}
  \centering
  \includegraphics[width=1\linewidth]{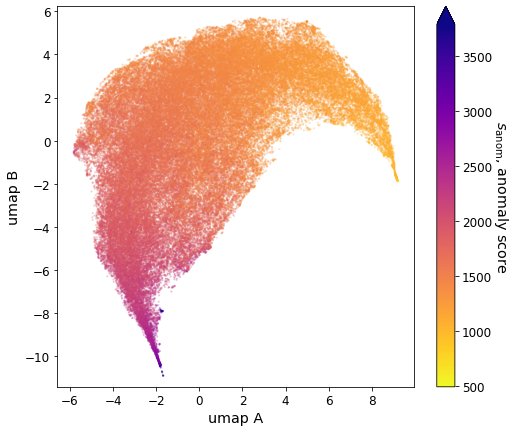}  
  \caption{}
  \label{fig:umap100k}
\end{subfigure}
\hfill
\vspace{0cm}
\begin{subfigure}{.495\textwidth}
  \centering
  \includegraphics[width=1\linewidth]{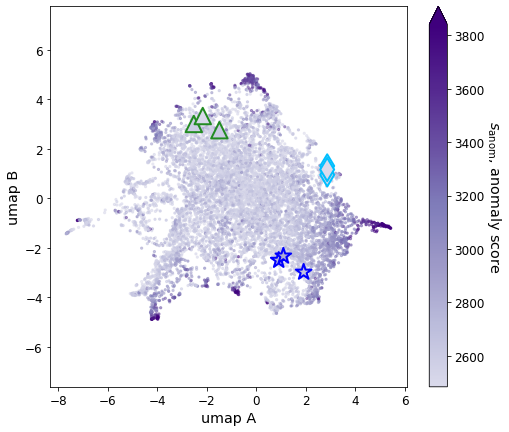}  
  \caption{}
  \label{fig:umap3sig}
\end{subfigure}
\begin{subfigure}{.32\textwidth}
  \centering
  \includegraphics[width=1\linewidth]{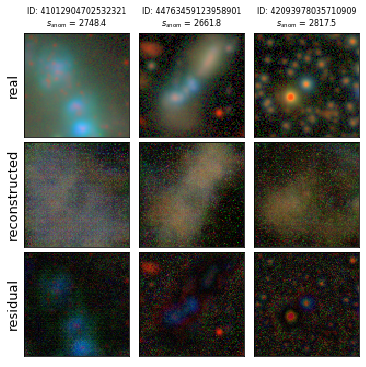}  
  \caption{}
  \label{fig:bluesf}
\end{subfigure}
\hfill
\begin{subfigure}{.32\textwidth}
  \centering
  \includegraphics[width=1\linewidth]{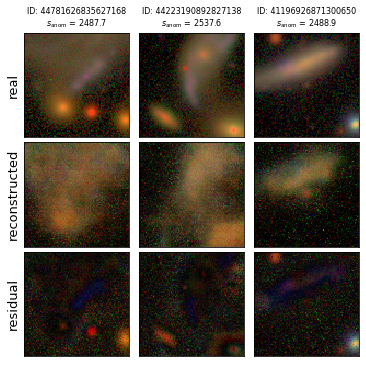}  
  \caption{}
  \label{fig:extendedpurple}
\end{subfigure}
\hfill
\begin{subfigure}{.32\textwidth}
  \centering
  \includegraphics[width=1\linewidth]{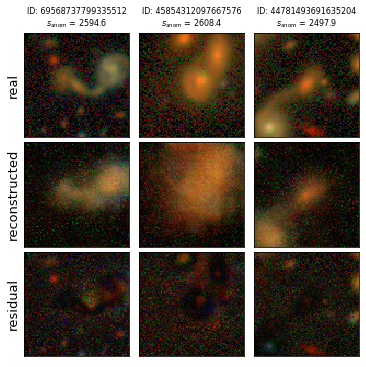}  
  \caption{}
  \label{fig:mergers}
\end{subfigure}
\vspace{0cm}
\caption{UMAPs of the autoencoded residuals between WGAN-reconstructed and real images, for (a) a random 100,000-object subsample and (b) objects with anomaly scores >3$\sigma$ above the mean. Similar objects cluster together in the UMAP; neighboring interesting anomalies are outlined in (b), and shown in (c)-(e). We find (c) blue star-forming regions (dark blue stars), (d) extended galaxies with active regions (light blue diamonds), and (e) galaxy mergers (green triangles).}
\label{fig:umap}
\end{figure}

We visualize these autoencoded residual images with a Uniform Manifold Approximation and Projection (UMAP, \citealt{McInnes2018}), which maps the objects into a 2D representation.
We first embed a 100,000 object subsample of our data set (Figure \ref{fig:umap100k}); the objects are colored by anomaly score.
This shows a clear gradient, with the high-scoring objects clustered at one side of the UMAP, confirming that the CAE is capturing information relevant to the anomalousness of the image.
We compare this to applying the CAE directly to the original images, to understand the importance of the WGAN residuals for anomaly detection, and as the residuals could be affected by uninteresting artifacts.
We find a less clear correlation between the anomaly score and the UMAP clustering on the autoencoded original images, compared to the autoencoded residuals.
We thus conclude that the residuals contain useful information about anomalies and use them for further characterization, though future work could explore incorporating the originals to improve the characterization.

We next perform a new UMAP embedding on only the >3$\sigma$ anomalies, as these are the ones we seek to further characterize; this is shown in Figure \ref{fig:umap3sig}.
Note that the particular mapping coordinates are not relevant, only the position of the objects within each embedding.
This mapping displays structure corresponding to anomaly score, with high-scoring objects clustering around the edges.
As these are all high-scoring objects, we expect that the UMAP on the autoencoded residuals will provide a natural method for sorting the anomalies; we explore this in the following section.
We note that UMAPs scale well with large samples, as evidenced by the map of the 100,000 object sample in Figure~\ref{fig:umap100k}, so this approach to characterization should also work well with larger selections of anomalies.

\subsection{Identified Anomalies}

We utilize our WGAN-based anomaly scores and autoencoded residual image UMAP of >3$\sigma$ anomalies to characterize the objects in our data set.
We built a custom visualization tool to interactively explore the UMAP.\footnote{\texttt{https://weirdgalaxi.es}}
We find that similar types of anomalies cluster together in this space.

Interesting anomalous objects from three of these regions are shown in the lower panels of Figure \ref{fig:umap}, along with the WGAN reconstruction and the residual between it and the original image.
We identified extremely blue, apparently highly star-forming sources (Figure \ref{fig:bluesf}); extended galaxies with purple regions that may indicate star-formation activity as well (Figure \ref{fig:extendedpurple}); and galaxy merger events, or at least overlapping galaxies along the line of sight (Figure \ref{fig:mergers}).
The residual images demonstrate the effectiveness of clustering in this space (aided by CAE dimensionality reduction); objects in each group display similar features that are not well-captured by the WGAN and thus stand out in the residuals, such as the clear purple streaks in Figure \ref{fig:extendedpurple}.
The proximity of these anomalies in the space of our UMAP indicates that our anomaly detection and characterization method is efficiently identifying objects with high scientific potential.

In order to confirm the scientific interest of these identified anomalies, we chose several seemingly interesting high-anomaly-scoring objects and conducted follow-up spectroscopic observations with the Keck telescope. 
One of these was a blue compact source overlapping a diffuse galaxy.
The analysis is currently in progress, but the spectrum shows that it is an extremely star-forming, metal-poor source at z$\sim$0.03 that may be a Blue Compact Dwarf galaxy, which are rare in the local universe.
It also displays asymmetrical emission features, potentially indicating gaseous outflows and solidifying the fact that our approach detected a scientifically interesting object.

Our method also proves useful for separating pipeline errors from interesting anomalies.
Similar types of errors, such as saturation in a color band or streaks across the image, cluster together in the autoencoded WGAN-residual UMAP.
This could be used to robustly filter out such errors.

\section{Conclusions}

In this work, we have shown that generative adversarial networks are a promising approach for anomaly detection in astronomical imaging.
We applied our method to $\sim$940,000 images of galaxies from the Hyper Suprime-Cam, and proposed a novel CAE-based approach to further characterize the anomalies.
We have identified and catalogued a set of interesting anomalies, and performed follow-up observations on several of these.
We plan to publicly release a catalog of our full data set with our WGAN-assigned anomaly scores, together with our custom visualization tool, which we hope will trigger new discoveries. 
Our WGAN + CAE + visualization method provides an approach to anomaly detection that is scalable, reproducible, and removes spontaneity from the discovery process, making it ideal for extracting novel science from the increasingly large surveys of the coming decade.

\section*{Broader Impact}

We hope the present work will help astronomers to make new discoveries in the era of big-data astronomy. 
The work uses public data of the deep sky acquired using a ground-based facility. 
We believe this work does not entail any negative consequences or ethical issues.

\begin{ack}
We gratefully acknowledge the Kavli Summer Program in Astrophysics for seeding this project; the initial work was completed at the 2019 program at the University of California, Santa Cruz.
This work was funded by the Kavli Foundation, the National Science Foundation, and UC Santa Cruz.
KSF thanks Dezso Ribli, Lorenzo Zanisi, Itamar Reis, and the Flatiron Astrodata Group at the Center for Computational Astrophysics for helpful discussions.
KSF, AL, and YL are grateful for valuable insights from Jenny Greene, Erin Kado-Fong, Kevin Bundy, Xavier Prochaska, Masami Ouchi, Kimihiko Nakajima, Yuki Isobe, Yi Xu, and Aaron Romanowsky.
NR's work at Argonne National Laboratory was supported under the U.S. Department of Energy contract DE-AC02-06CH11357. 
\end{ack}

\bibliographystyle{mnras}
\bibliography{Anomalies-HSC}

\end{document}